\documentclass[aps,prb,twocolumn,superscriptaddress,floatfix,10pt]{revtex4-1}

\usepackage{graphicx}
\usepackage{amsmath}
\usepackage{amssymb}
\usepackage[bookmarksnumbered=true,hypertexnames=false,colorlinks=true,linkcolor=blue,urlcolor=blue,citecolor=blue,anchorcolor=green]{hyperref}

\newcommand{\ket}[1]{\mbox{$| #1 \rangle$}}
\newcommand{\braket}[2]{\mbox{$\langle #1 | #2 \rangle$}}

\newcommand{\tr}{\mbox{$\text{tr}$}}

\begin{document}

\title{Loop Series Expansions for Tensor Networks}
\author{Glen Evenbly}
\email{evenbly@amazon.com}
\author{Nicola Pancotti}
\author{Ashley Milsted}
\affiliation{AWS Center for Quantum Computing, Pasadena, CA 91125, USA}
\affiliation{California Institute of Technology, Pasadena, CA 91125, USA}
\author{Johnnie Gray}
\author{Garnet Kin-Lic Chan}
\affiliation{California Institute of Technology, Pasadena, CA 91125, USA}
\date{\today}

\begin{abstract}
Belief propagation (BP) can be a useful tool to approximately contract a tensor network, provided that the contributions from any closed loops in the network are sufficiently weak. In this manuscript we describe how a loop series expansion can be applied to systematically improve the accuracy of a BP approximation to a tensor network contraction, in principle converging arbitrarily close to the exact result. More generally, our result provides a framework for expanding a tensor network as a sum of component networks in a hierarchy of increasing complexity.  We benchmark this proposal for the contraction of iPEPS, either representing the ground state of an AKLT model or with randomly defined tensors, where it is shown to improve in accuracy over standard BP by several orders of magnitude whilst incurring only a minor increase in computational cost. These results indicate that the proposed series expansions could be a useful tool to accurately evaluate tensor networks in cases that otherwise exceed the limits of established contraction routines.
\end{abstract}

\maketitle

%%%%%%%%%%%%%%%%%%%%%%%%%%%%%%%%%%%%%%%%%%%%%%%%%%%%%%%%%%%%%%%%%%%%%%%%%
\section{Introduction} \label{sect:intro}
Tensor networks (TNs) are powerful tools for the classical simulation of quantum systems \cite{TN1,TN2,TN3,TN4,TN5,TN6,TN7,TN8,TN9,TN10}. Although they were traditionally developed for model systems in the context of condensed matter physics, they have more recently been applied towards the benchmarking of nascent quantum computers \cite{QC1,QC2,QC3,QC4,QC5,QC6,QC7,QC8,QC9,QC10,QC11,QC12}. Tensor network states were pioneered by the introduction of matrix product states \cite{MPS1,MPS2} (MPS), which possess tensors connected in a $1D$ geometry. For several decades numeric algorithms based on the MPS, such as the density matrix renormalization group\cite{DMRG1,DMRG2,DMRG3} (DMRG), have been widely regarded as the gold-standard of classical simulation algorithms for $1D$ quantum systems; producing high-accuracy and unbiased results for a wide range of problems. A large part of this success can be attributed to the computational efficiency with which MPS can be contracted. Subsequent developments have seen the proposal of sophisticated tensor networks, such as projected entangled pair states\cite{PEPS1, PEPS2} (PEPS) and multi-scale entanglement renormalization ansatz\cite{MERA1, MERA2} (MERA), designed to efficiently represent quantum states on large $2D$ systems. However, a significant challenge of these $2D$ tensor networks is the high computational expense required for their contraction \cite{PEPS3} even when approximate contraction schemes are used \cite{PEPS4, PEPS5, PEPS6, PEPS7, PEPS7b, PEPS7c, PEPS8}. In practice, this expense limits the amount of entanglement that can be encoded in these networks and therefore also limits the range of problems for which they produce accurate results. Thus the development of strategies for more efficient contraction of TNs remains a vital task.

A promising approach for the contraction of TNs is through the adoption of belief propagation\cite{BP0,BP1,BP2} (BP), a message passing algorithm with a long history in both computer science\cite{BP3,BP4,BP5} and in statistical physics\cite{BP6,BP7,BP8}. In recent times the application of BP towards TNs has generated much interest \cite{BP9,BP10,BP11,BP12,BP13,BP14,BP15,BP16,BP17,BP18}, and produced a notable success \cite{QC11} of simulating a quantum experiment performed on IBMs 127-qubit Eagle processor \cite{IBM1}, a problem that was seen to be intractable for other $2D$ TN algorithms \cite{ISO1}. However, a major drawback of BP is that it is an \emph{uncontrolled approximation}; one cannot systematically improve the accuracy of a BP result as is otherwise done in the context of tensor networks by simply growing the bond dimension, although there do exist schemes to improve the accuracy of BP through a preliminary clustering of a network \cite{BP15}. 

In this manuscript we present a method to improve upon the accuracy of BP for the contraction of a tensor network by incorporating and building upon the loop series expansion proposed by Chertkov and Chernyak\cite{LOOP1,LOOP2}. Assuming that a BP fixed point of the tensor network has been found, we describe how the dominant loop excitations, otherwise neglected by BP, can be re-incorporated into the evaluation of this network. This approach can also be understood as providing a general notion of a \emph{series expansion for tensor networks}; that a complicated tensor network may be expanded as a sum of component networks in a hierarchy of increasing complexity, with the BP approximation as the zeroth-order term in the series. We demonstrate numerically that the proposed series expansion can provide both accurate and computationally cheap evaluations of PEPS, such that it could be a useful tool to contract tensor networks in cases that otherwise exceed the limits of established contraction routines.

%%%%%%%%%%%%%%%%%%%%%%%%%%%%%%%%%%%%%%%%%%%%%%%%%%%%%%%%%%%%%%%%%%%%%%%%%
\section{Belief Propagation} \label{sect:BP}
Here we provide a brief introduction to BP as a necessary precursor to our main results. Let $\mathcal T$ be a closed tensor network defined of $N$ tensors labelled $\{T_1, T_2, T_3, \ldots, T_N \}$ and with $M$ indices labelled $\{i_1, i_2, i_3, \ldots i_M \}$. Assume that our goal is to contract $\mathcal T$ to a scalar $\mathcal{Z}(\mathcal{T})$,
\begin{equation}
\mathcal{Z}(\mathcal{T}) = \sum_{i_1,i_2,\ldots} T_1 (\xi_1) \ T_2 (\xi_2) \ T_3 (\xi_3) \ldots, \label{eq:sum}
\end{equation}
where $\xi_r$ describes the subset of indices $\xi_r \subseteq \{i_1, i_2, \ldots i_M \}$ connected to tensor $T_r$. The starting point of BP is to define a pair of messages for each edge of the network; if $(r,s)$ denotes an edge shared between tensors $T_r$ and $T_s$ then we define messages $\{\mu_{r\to s}, \mu_{s\to r} \}$ on this edge. We refer to a message $\mu_{s\to r}$ as outgoing from $T_s$ and incoming to $T_r$. BP prescribes the following algorithm to update each of the messages: the new outgoing message for $T_r$ towards tensor $T_s$ is given by contracting the incoming messages to $T_r$ on all other indices,
\begin{equation}
\mu_{r\to s} = \left( \prod_{q \in \xi_r / s} \mu_{q\to r} \right) T_r. \label{eq:BPupdate}
\end{equation}
The set of $2M$ messages realize a \emph{BP fixed point} if all of the equations are simultaneously satisfied up to multiplicative constants. Let us assume that we have found a BP fixed point for our network $\mathcal T$ and that each matching pair of messages have been normalized such that their dot product is unity, $\left\langle \mu_{r\to s}\cdot \mu_{s\to r} \right\rangle = 1$. Let us define scalar $\tilde T_r$ as resulting from the contraction of a tensor $T_r$ with all incoming messages,
\begin{equation}
\tilde T_r = \left( \prod_{q \in \xi_r} \mu_{q\to r} \right) T_r, \label{eq:BPvacuum}
\end{equation}
which we refer to as the (BP) vacuum contribution from tensor $T_r$. The BP approximation to the network contraction $\mathcal{Z}$ from Eq.~\ref{eq:sum} is given by multiplication of these scalars
\begin{equation}
\mathcal{Z}(\mathcal{T}) \approx \prod_{r} \tilde T_r. \label{eq:BPapprox}
\end{equation}
It follows that the free energy $\mathcal{F} = -\log \mathcal{Z}$ can be approximated as
\begin{equation}
    \mathcal{F}(\mathcal{T}) \approx - \sum_r \log \left( \tilde{T}_r \right), \label{eq:bethefree}
\end{equation}
which is also known as the Bethe free energy \cite{BP0}.

%%%%%%%%%%%%%%%%%%%%%%%%%%%%%%%%%%%%%%%%%
\begin{figure} [!th] %[!t!b]
\begin{center}
\includegraphics[width=7.0cm]{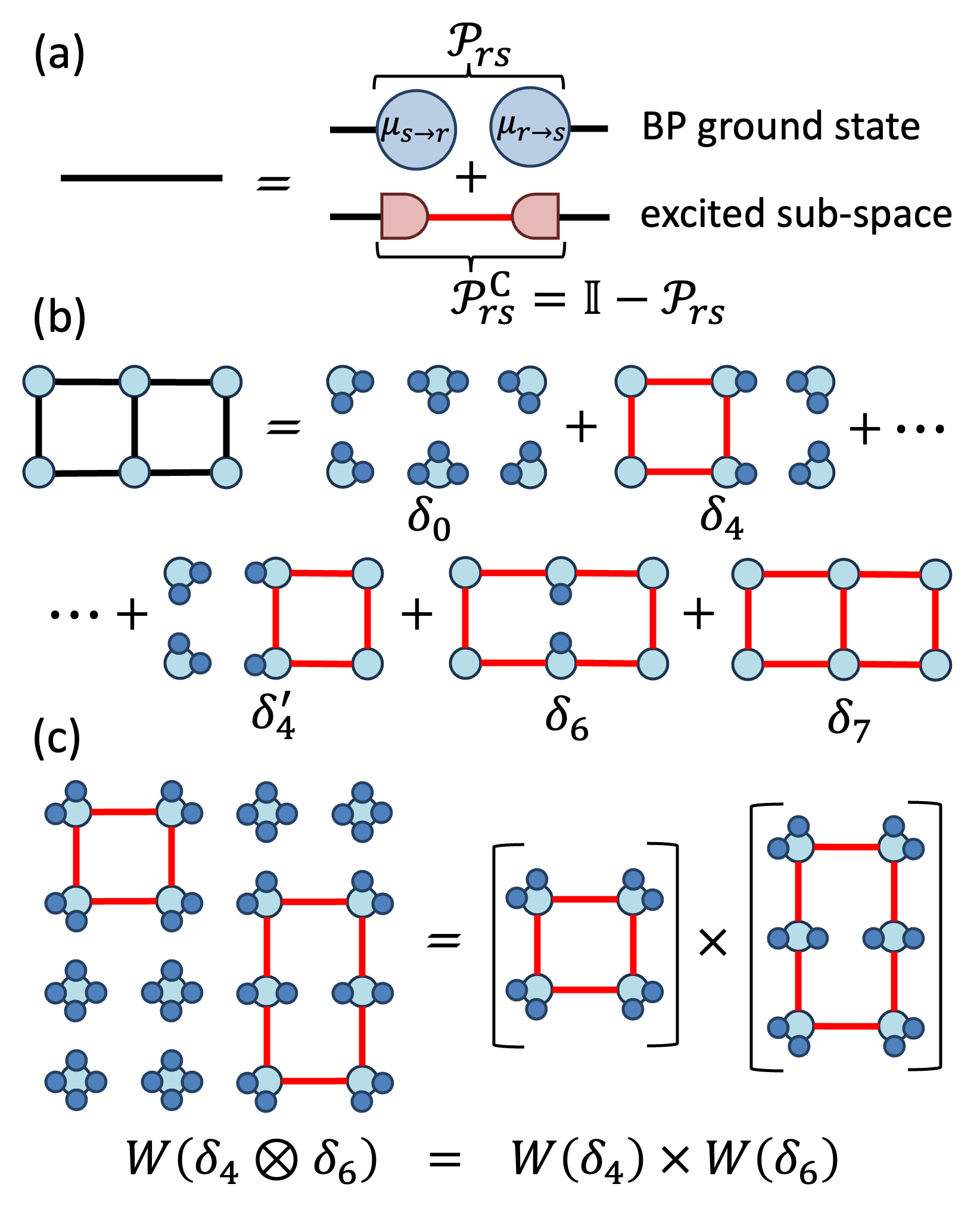}
\caption{(a) An edge within a tensor network is resolved into the sum of a projector onto the BP ground-state, formed from the outer product of the BP messages, and the complimentary projector onto the excited subspace. (b) A network is expanded as a sum of 5 configurations, where indices in each configuration have been projected onto either the ground or excited sub-space. The zero-weight configurations (i.e.\ those containing dangling excitations) have been omitted. (c) The weight of a configuration containing several distinct excitations is equal to the product of the individual excitation weights (assuming that the vacuum contributions defined in Eq.~\ref{eq:BPvacuum} have been normalized to unity).}
\label{fig:BP}
\end{center}
\end{figure}
%%%%%%%%%%%%%%%%%%%%%%%%%%%%%%%%%%%%%%%%%

%%%%%%%%%%%%%%%%%%%%%%%%%%%%%%%%%%%%%%%%%%%%%%%%%%%%%%%%%%%%%%%%%%%%%%%%%
\section{Configurations in the BP Basis} \label{sect:config}
We now outline a methodology to improve upon the accuracy of a known BP fixed point by introducing additional terms into Eq.~\ref{eq:BPapprox}. For any edge $(r,s)$ of the network, assumed to be of dimension $d$, we define a rank $1$ projector $\mathcal P_{rs}$ onto message subspace and its rank $(d-1)$ compliment $\mathcal P_{rs}^\textrm{C}$,
\begin{equation}
\mathcal P_{rs} = \mu_{s\to r} \otimes \mu_{r\to s},\ \ \ \ \ \mathcal P_{rs}^\textrm{C} = \mathbb{I} - \mathcal P_{rs}. \label{eq:proj}
\end{equation}
with $\mathbb{I}$ as the $d$-dim identity, see also Fig.~\ref{fig:BP}(a).
We henceforth refer to $\mathcal P_{rs}$ as the projector onto the (BP) ground state of edge $(r,s)$ and $\mathcal P_{rs}^\textrm{C}$ as the projector onto the excited subspace. It follows that the tensor network $\mathcal T$ can be resolved as a sum over $2^M$ (BP basis) configurations, formed from all combinations of projecting each edge into either the ground or excited sub-space. We define the \emph{degree} of a configuration as the number of excited edges that it contains, using $\delta_x$ to denote a degree-$x$ configuration and $W(\delta_x)$ to denote its \emph{weight} (i.e.\ the scalar resulting from its contraction). Thus we can recognize $\delta_0$ as the \emph{BP vacuum}; the unique configuration with all edges projected into their BP ground state. It follows that $W(\delta_0)$ evaluates to the standard BP result of Eq.~\ref{eq:BPapprox}. Our goal is now to characterize the weights of the remaining configurations, which may be thought of as excitations on top of the BP vacuum, such that the high-weight contributions can then be re-introduced into the approximation. 

A significant step towards this goal comes from the understanding derived in Ref.~\onlinecite{LOOP1} that any configuration with a \emph{dangling excitation}, i.e.\ that contains a tensor with a single excited index, has weight zero. This result follows directly from the definition of the BP fixed point; let us assume that edge $(r,s)$ of tensor $T_r$ is projected via $\mathcal P^\textrm{C}_{rs}$ into the excited subspace and all other edges are in the BP ground. It follows that
\begin{equation}
 \left( \prod_{q \in \xi_r / s} \mu_{q\to r} \right) T_r  \mathcal P^\textrm{C}_{rs} = \mu_{r\to s} \mathcal P^\textrm{C}_{rs} = 0, \label{eq:dangling}
\end{equation}
where Eq.~\ref{eq:BPupdate} was used to replace $T_r$ with the outgoing message $\mu_{r\to s}$ which is, by construction in Eq.~\ref{eq:proj}, orthogonal to the projector $\mathcal P^\textrm{C}_{rs}$. This result, which implies that non-trivial excitations can only occur in closed loops, is useful in reducing the number of basis configurations that need be considered. For instance, when applied to the network of $M=7$ edges in Fig.~\ref{fig:BP}(b), the number of configurations is reduced from $2^7=128$ to only $5$ configurations of non-zero weight labelled $\{\delta_0, \delta_4, \delta_4', \delta_6, \delta_7 \}$. Another useful observation, as examined in Ref.~\onlinecite{LOOP3}, is that the weight of a configuration containing separate excitations, i.e.\ excitations that do not share any common tensors, is given by the product of the individual weights
\begin{equation}
W(\delta_x \otimes \delta_{y}) = W(\delta_x) \times W(\delta_{y}),\label{eq:product}
\end{equation}
see also the example shown in Fig.~\ref{fig:BP}(c). This result is further simplified by rescaling each tensor $T_r$ so that its vacuum state contribution, i.e. $\tilde T_r$ from Eq.~\ref{eq:BPvacuum}, is unity which allows the vacuum component of any configuration to be ignored. This observation implies that one only needs to understand the connected excitations in order to fully characterize a tensor network contraction.

For any reasonably large tensor network the full number of non-trivial excitations will still exceed practical consideration. However, as detailed further in Appendix~\ref{sect:degree}, we posit that in most spatially homogeneous tensor networks the weight of a configuration $\delta_x$ will be exponentially suppressed in its degree $x$, 
\begin{equation}
W(\delta_x) \approx \exp (-k x)\label{eq:suppression}
\end{equation}
with $k$ a positive scalar. While it is easy to construct counterexamples where this scaling will not hold, it is argued in Appendix~\ref{sect:degree} that this behavior will hold for networks where the BP fixed point already provides a reasonably accurate approximation to the network contraction. Thus one can understand that the (BP basis) configurations, when ordered in increasing degree, form a series expansion for a network contraction. While the number of $\delta_x$ configurations may still grow (exponentially) quickly with $x$, the exponential suppression of the contribution of each configuration provides the possibility that a partial summation will rapidly converge to the exact result as higher-degree excitations are added. 

\section{Contraction of iPEPS} \label{sect:peps}
\begin{figure} [!t] %[!t!b]
\begin{center}
\includegraphics[width=8.0cm]{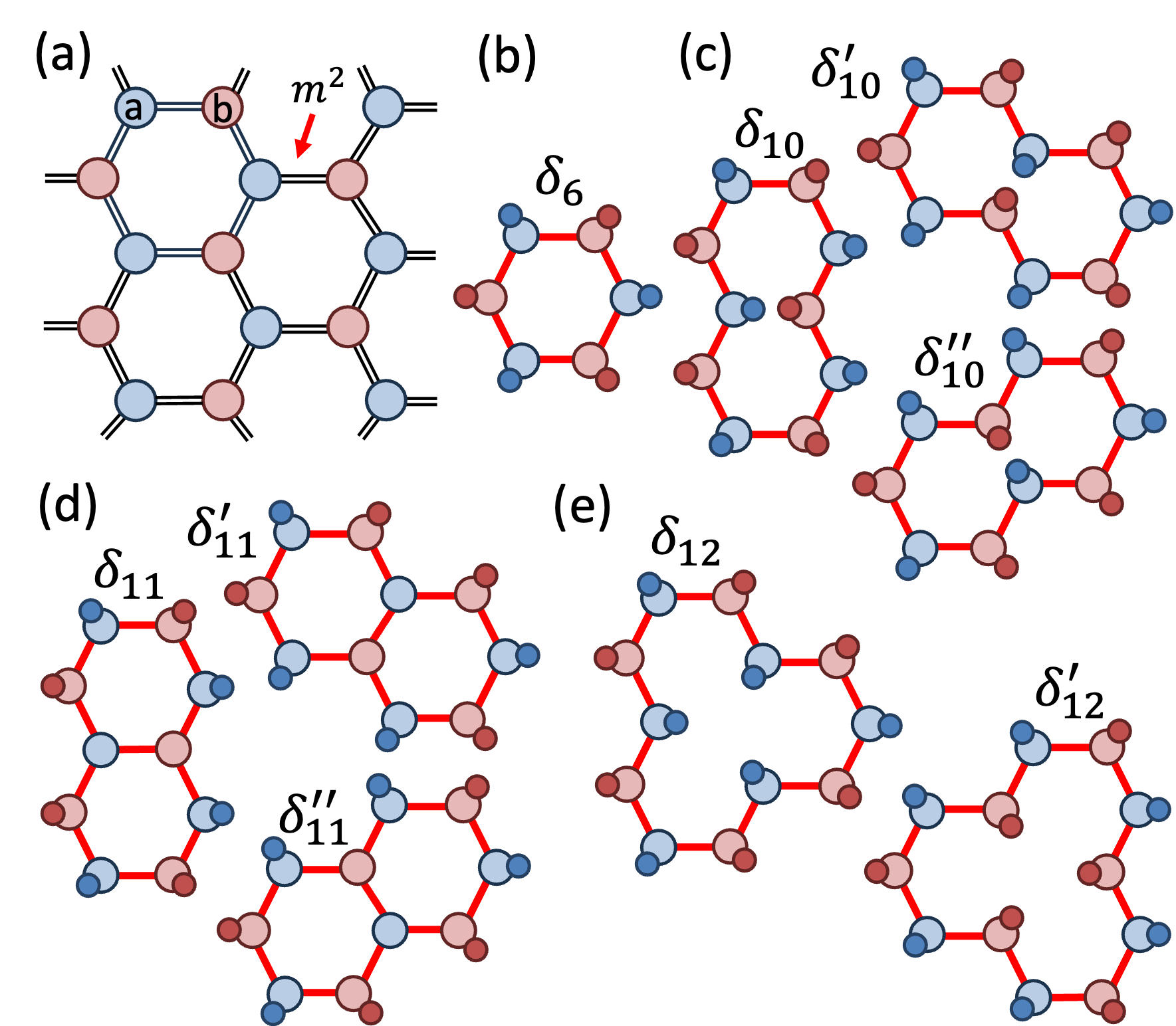}
\caption{(a) A closed tensor network representing $\mathcal T = \braket{\psi}{\psi}$ results from contracting the physical indices of a hexagonal lattice PEPS $\ket{\psi}$, bond dimension $m$, with its conjugate. (b-e) Depiction of the excitations $\delta_x$ up to $12^\textrm{th}$ degree.}
\label{fig:free}
\end{center}
\end{figure}

In this section we discuss how the proposed series expansion can be applied to contract infinite projected entangled pair states (iPEPS). For concreteness we focus on a hexagonal lattice PEPS $\ket{\psi}$, local dimension $d$ and virtual dimension $m$, formed from a single pair of unique tensors $A$ and $B$ tiled on alternate sites (such that the state has a 2-site unit cell). A closed hexagonal lattice tensor network $\mathcal T = \braket{\psi}{\psi}$ of bond dimension $m^2$ is formed by taking the product of the PEPS with its conjugate and contracting over the site indices, see also Fig.~\ref{fig:free}(a). The network $\mathcal T$ is composed of two unique tensors $a$ and $b$, which result from contraction of the physical index in $(A\otimes A^*)$ and $(B\otimes B^*)$ respectively. In what follows we assume that we have found a BP fixed point for $\mathcal T$ and that tensors $a$ and $b$ have been re-scaled such that their vacuum contributions, as defined in Eq.~\ref{eq:BPvacuum}, are unity.

%%%%%%%%%%%%%%%%%%%%%%%%%%%%%%%%%%%%%%%%%
%%%%%%%%%%%%%%%%%%%%%%%%%%%%%%%%%%%%%%%%%

\subsection{Free-energy density} \label{sect:free}
A common task in tensor network algorithms is to evaluate the scalar $\mathcal Z$ associated to contracting a closed tensor network, as per Eq.~\ref{eq:sum}. However in the context of an infinite tensor network this scalar would diverge so we instead evaluate an intensive equivalent: the free-energy density $f$,
\begin{equation}
f = \lim_{N \to \infty} \left( \frac{-\textrm{log}(\mathcal Z_N)}{N} \right). \label{eq:free}
\end{equation}
where $\mathcal Z_N$ represents the evaluation of a patch of $N$ tensors from the full network. Note that, due to our normalization of the BP vacuum, the BP contribution to $\mathcal Z_N$ is unity for all $N$ (or equivalently the Bethe free energy is zero \cite{BP0}).

Our first step is to evaluate the weights $W(\delta_x)$ of the lowest degree excitations $\delta_x$; in the calculations performed in this manuscript we evaluate the excitations up to degree $x=14$, see Fig.~\ref{fig:free}(b-e). Each of these weights is given by contracting a small tensor network formed by capping the external indices (i.e. of tensors within the excitation) with their fixed point messages and projecting the internal indices onto their excited subspace with projectors $\mathcal P_{rs}^\textrm{C}$ as defined in Eq.~\ref{eq:proj}. However, even once these weights $W(\delta_x)$ are known, it still remains a non-trivial combinatorial task to compute their impact on the free energy density $f$. Here we shall follow the approximate counting strategy derived in Appendix~\ref{sect:counting}.2, which gives the following result for the free energy correction
\begin{equation}
f \approx -\sum_{\{\delta_x\}} L(\delta_x) W(\delta_x) e^{S(\delta_x) f}. \label{eq:f2}
\end{equation}
with $L(\delta_x)$ as the number of locations per lattice site that excitation can be placed and $S(\delta_x)$ as the number of hexagonal plaquettes that the excitation occupies. For instance, the $\delta_6$ excitation shown in Fig.~\ref{fig:BP}(b) has $L(\delta_6) = 1/2$ (since there is one hexagon per two lattice sites) and $S(\delta_6) = 7$. Given that Eq.~\ref{eq:f2} contains the unknown $f$ on both sides we cannot solve for $f$ directly. Instead we solve Eq.~\ref{eq:f2} iteratively by first setting $f=0$ on the r.h.s, computing a new value for $f$, then feeding the new $f$ back into r.h.s and repeating until $f$ is converged.

% Usually this requires only a few iterations to converge to high precision. 
% Notice that the terms $\exp(S(\delta_x) f)$ in Eq.~\ref{eq:f2} act to suppress excitations based on the amount of space they occupy.

\subsection{Transfer matrix} \label{sect:transfer}
\begin{figure} [!t] %[!t!b]
\begin{center}
\includegraphics[width=8.0cm]{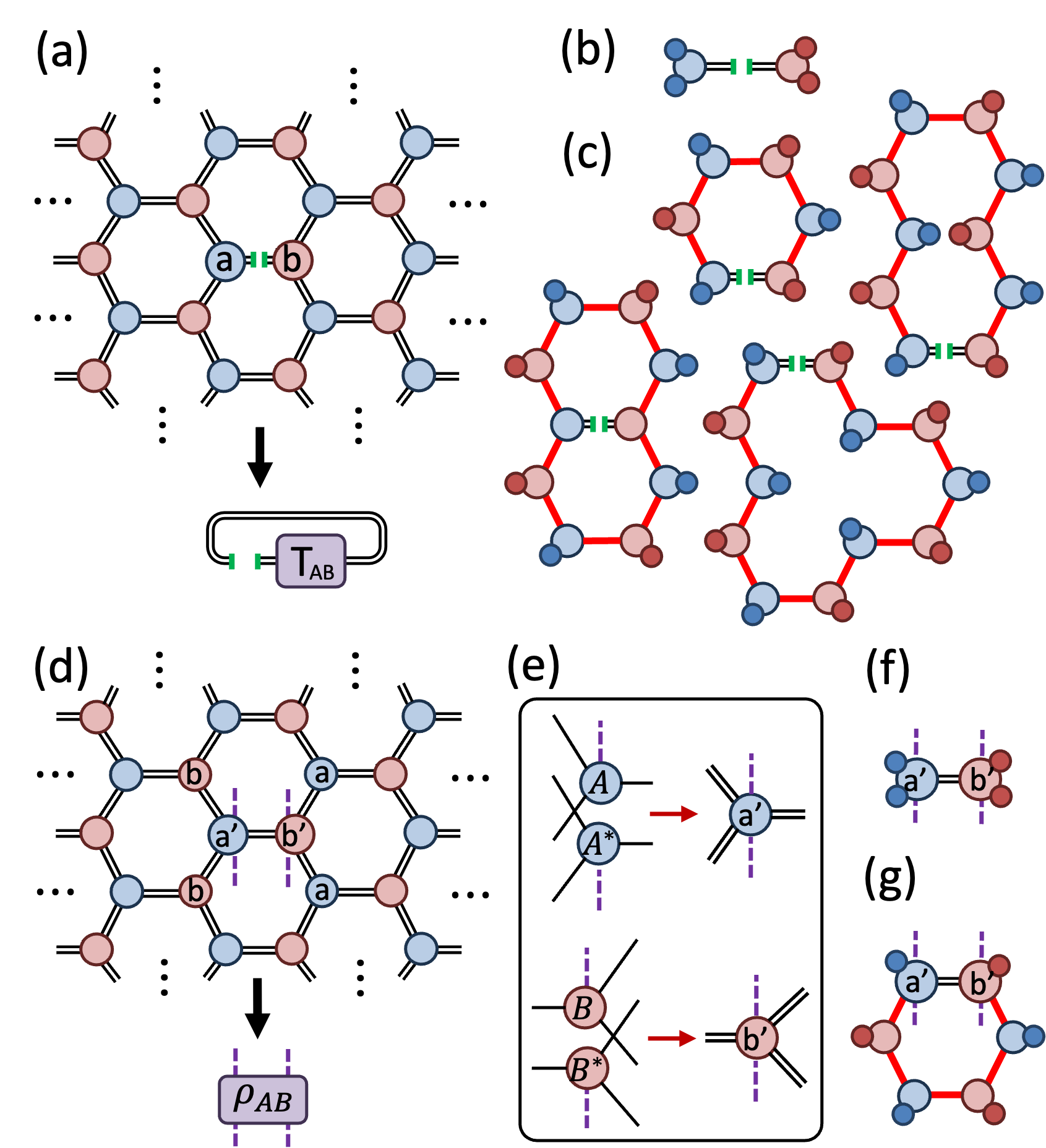}
\caption{(a) The transfer matrix $\textrm{T}_{AB}$ is formed by contracting the network after a single edge has been cut. (b) The BP vacuum contribution to $\textrm{T}_{AB}$. (c) Several examples of loop corrections to $\textrm{T}_{AB}$. (d) The density matrix $\rho_{AB}$ is obtained from contracting the network with a pair of impurity tensors $a'$ and $b'$ that have their physical indices left open. (e) Definition of the impurity tensors $a'$ and $b'$. (f) The BP vacuum contribution to $\rho_{AB}$. (g) A loop correction to $\rho_{AB}$.}
\label{fig:transfer}
\end{center}
\end{figure}

We now examine how one can compute the transfer matrix $\textrm{T}_{AB}$ associated to cutting open a single edge from the network $\mathcal T = \braket{\psi}{\psi}$ as shown in Fig.~\ref{fig:transfer}(a). Such transfer matrices are utilized in determining the optimal truncation of an internal index from the PEPS and thus play an important role in many PEPS optimization algorithms \cite{PEPS4,PEPS5,PEPS6,PEPS7,PEPS7b,PEPS7c}.

The BP vacuum contribution to $\textrm{T}_{AB}$ is shown in Fig.~\ref{fig:transfer}(b) together in Fig.~\ref{fig:transfer}(c) with examples of contributions from loop excitations, all of which individually evaluate to a matrix with the same dimensions as $\textrm{T}_{AB}$. The series approximation to $\textrm{T}_{AB}$ is constructed by evaluating each of the loop-excitation networks (up to some desired degree), then adding the contributions together after weighting each by a suppression factor $\exp(S(\delta_x)f)$ similar to Eq.~\ref{eq:f2}. Note this requires that the free energy density $f$ be evaluated prior to computing $\textrm{T}_{AB}$. Finally, we remark that the excitation terms that do not include either of the tensors adjacent to an open index do not need to be considered explicitly; their contributions, which are proportionate to the vacuum term, are already accounted for through the suppression factors. 

%%%%%%%%%%%%%%%%%%%%%%%%%%%%%%%%%%%%%%%%%
%%%%%%%%%%%%%%%%%%%%%%%%%%%%%%%%%%%%%%%%%

\subsection{Density matrix} \label{sect:density}
We now describe how the density matrix $\rho_{AB}$ associated to a pair of adjacent lattice sites can be computed. Let us first define tensor $a'$ from taking $(A\otimes A^*)$ while leaving the physical indices uncontracted and similarly for $b'$, see Fig.~\ref{fig:transfer}(e). The network for $\rho_{AB}$ can be realized from a substitution of $a'$ and $b'$ as impurities in the original network $\mathcal T = \braket{\psi}{\psi}$ as shown in Fig.~\ref{fig:transfer}(d).

Similarly to the transfer matrix, only the loop excitations that include one (or both) of the tensors $a'$ and $b'$ will give non-trivial contribution to $\rho_{AB}$. The BP vacuum contribution to $\rho_{AB}$ is shown in Fig.~\ref{fig:transfer}(f) together in Fig.~\ref{fig:transfer}(g) with an example of a loop excitation. The series approximation to $\rho_{AB}$ is thus constructed in the same way as the transfer matrix; the contributing loop-excitation networks are evaluated up to some desired degree and then added together after weighting each by a suppression factor $\exp(S(\delta_x)f)$. 

\section{Benchmark results} \label{sect:benchmark}
\begin{figure} [!t] %[!t!b]
\begin{center}
\includegraphics[width=8.5cm]{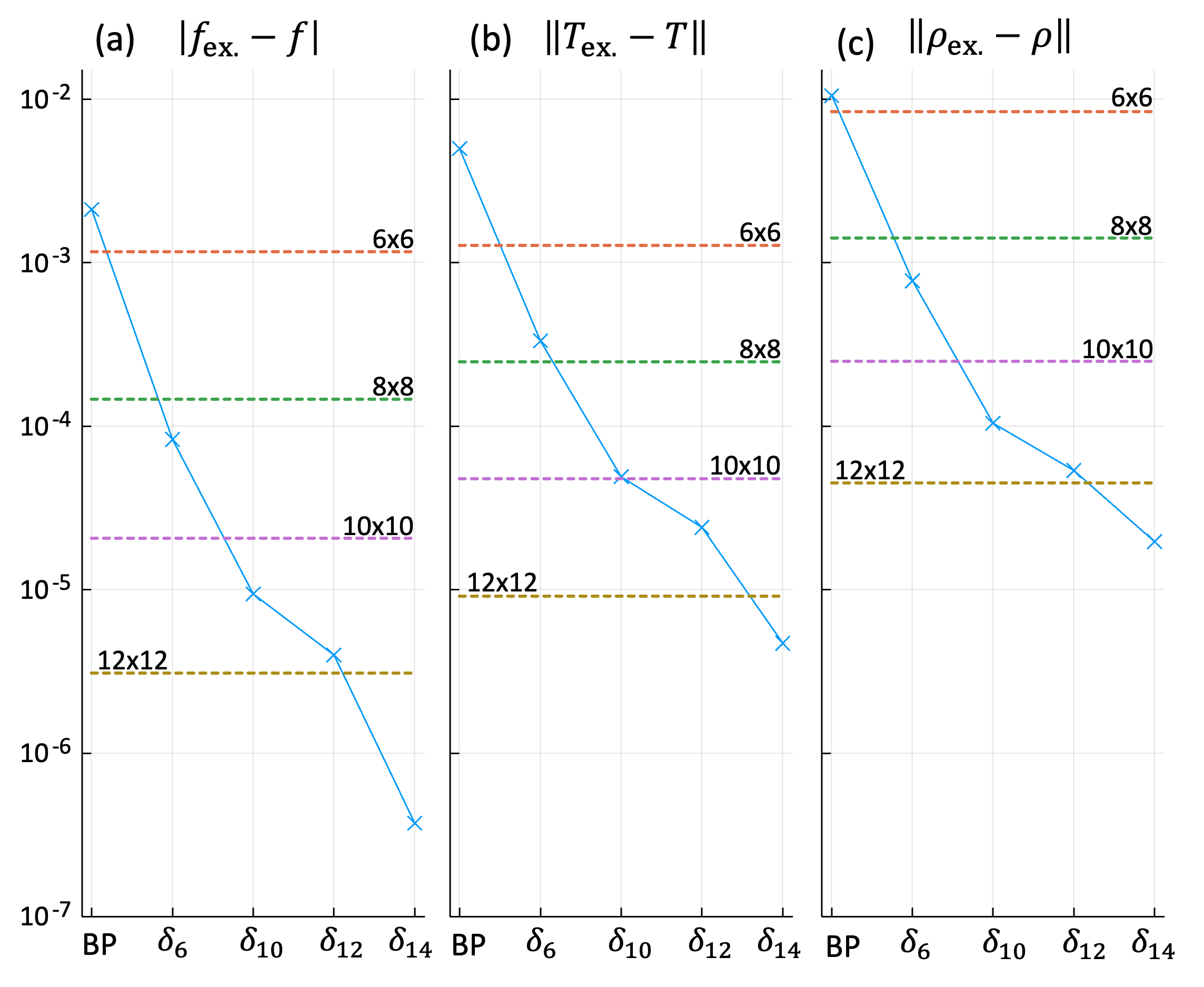}
\caption{Results comparing the accuracy of BP and loop expansion up to degree $\delta_x$ against exact results for the ground state of the (thermodynamic limit) hexagonal AKLT model. Displayed are errors in (a) the free-energy density $f$, (b) the transfer matrix $\textrm{T}_{AB}$, (c) the two-site density matrix $\rho_{AB}$. Dashed lines show results from an exact contraction of an $N\times N$ system.}
\label{fig:hexresults}
\end{center}
\end{figure}

We now benchmark our approach, beginning with contraction of a PEPS representing the ground state $\ket{\psi}$ of the hexagonal lattice AKLT model\cite{AKLT1,AKLT2} in the thermodynamic limit. The AKLT model is particular useful as a testbed, since $\ket{\psi}$ can be represented exactly\cite{PEPS1} with an iPEPS of bond dimension $m=2$. Following the methodology prescribed in Sect.~\ref{sect:peps} we compute the free energy density $f$, transfer matrix $\textrm{T}_{AB}$, and the density matrix $\rho_{AB}$ for loop expansions containing up to $14^\textrm{th}$ degree excitations $\delta_{14}$. In all cases we compare against evaluations using a boundary MPS contraction\cite{PEPS5} whose dimension $\chi$ has been chosen large enough, $\chi\approx 30$, such that any truncation errors are negligible and the results can be considered as numerically exact. Fig.~\ref{fig:hexresults} shows that the addition of the loop corrections improve the accuracy of the BP results dramatically, such that they appear to converge towards the exact results as the degree of the expansion is increased. The expansion to $14^\textrm{th}$ degree excitations $\delta_{14}$ gave the free energy density $f$ accurate to $\varepsilon = 3\times10^{-7}$, which represents an improvement of four orders of magnitude over the BP result. That the infinite hexagonal tensor network can be characterised so accurately from only the BP vacuum plus a few of the low-degree loop excitations is remarkable. Note that we also compare against results obtained from the exact contraction of the PEPS on a finite $N\times N$ geometry (with periodic boundaries) for $N=\{6,8,10,12 \}$. In all cases we see that an $N^\textrm{th}$ degree loop expansion roughly approximates the accuracy of an exact evaluation on a $N\times N$ geometry, providing additional confirmation that the loop expansions are properly accounting for all short-range contributions. 

% This is expected since the $N\times N$ systems would contain degree-$N$ corrections (i.e.\ that `wrap' the whole system) whereas a degree-$N$ loop expansion is only neglecting excitations of degree larger than $N$. None-the-less, this provides additional confirmation that the loop expansions are properly accounting for all short-range contributions.

\begin{figure} [!ht] %[!t!b]
\begin{center}
\includegraphics[width=8.5cm]{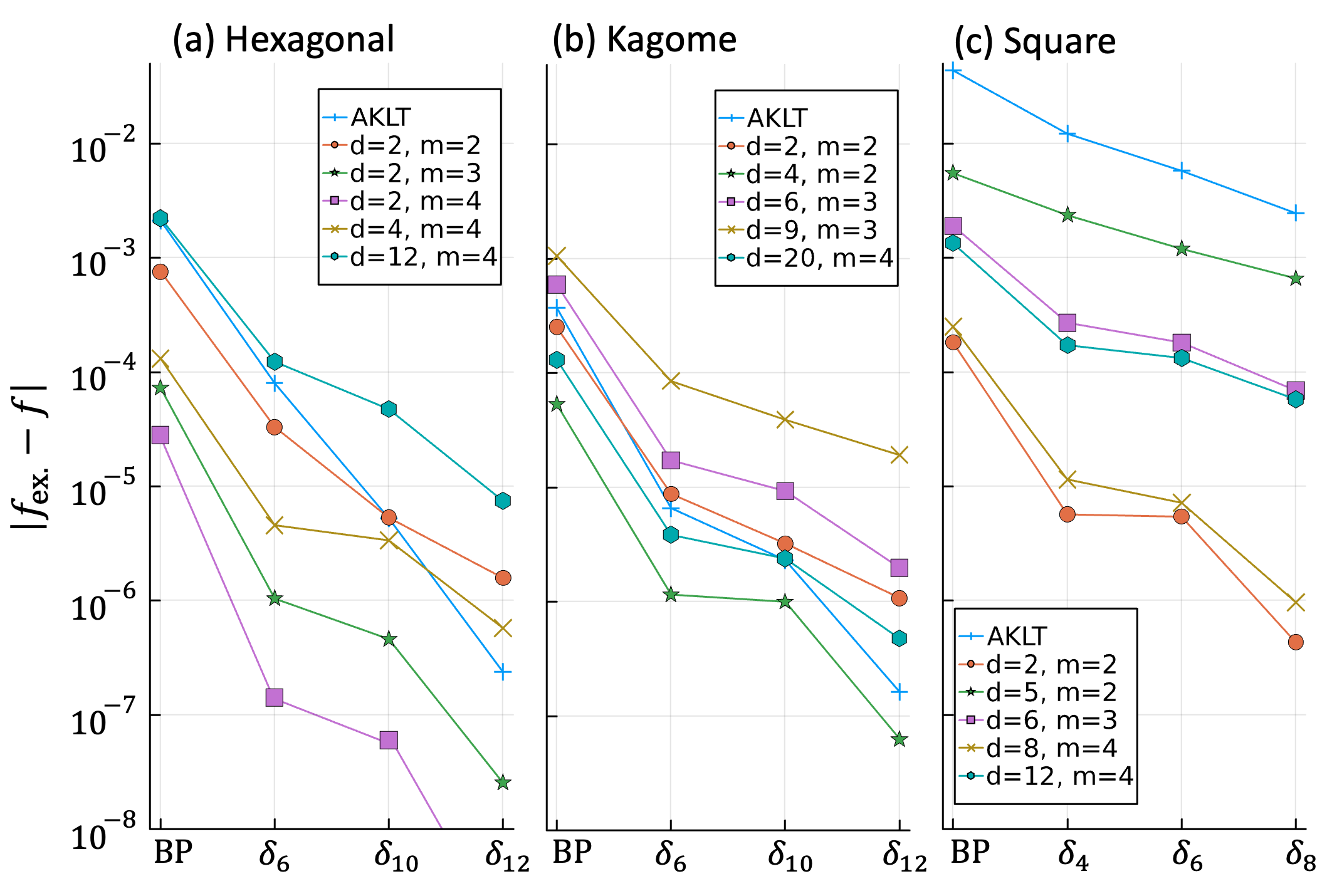}
\caption{Results comparing the accuracy of BP and loop expansion up to degree $\delta_x$ for evaluation of the free-energy density $f$ from iPEPS on either (a) the hexagonal lattice, (b) the kagome lattice, or (c) the square lattice. The iPEPS either encode the AKLT ground state (on each respective lattice geometry) or are formed from random tensors with local dimension $d$ and bond dimension $m$.}
\label{fig:allresults}
\end{center}
\end{figure}

In Fig.~\ref{fig:allresults}(a) we present results for the accuracy of the free-energy density evaluated from randomly defined iPEPS (with 2-site unit cell) of local dimension $d$ and bond dimension $m$. These were prepared by placing on each edge of the lattice a maximally entangled pair of dim-$m$ particles and then projecting each vertex onto a dim-$d$ space via a $(m^3\times d)$ isometry. The isometries were formed by first populating a $(m^3\times d)$ matrix with random elements chosen uniformly from the interval $[0,1]$, then enforcing rotational symmetry (useful to reduce the number of unique terms in the loop expansion), and finally orthogonalizing the columns of the matrix. The results from random iPEPS, which are broadly similar to those from the AKLT model, demonstrate that the series expansion seems to be valid for typical instances of iPEPS. Notice that increasing the local dimension $d$ (while keeping a fixed bond dimension $m$) was generally found to reduce the accuracy of our approach; this could be expected as increasing $d$ also increases the per-site entanglement entropy of the random PEPS.

Fig.~\ref{fig:allresults}(b-c) presents equivalent results for kagome and square lattices. For iPEPS defined on kagome lattices we perform local manipulations to exactly transform the network into a hexagonal geometry (similar to the form of PEPS described in Ref.~\onlinecite{PESS1}) prior to applying the series expansion, see Appendix \ref{sect:kagome} for details. The square lattice results were obtained by applying the expansion directly to the square lattice, evaluated up to the $\delta_8$ excitations as shown in Appendix \ref{sect:kagome}. Similar to the hexagonal lattice, the results from the kagome and square lattices also show robust improvements over the BP approximation. However, we see that the convergence on the square lattice is slower with respect to increasing the degree of expansion; we attribute this to the fact that the quantity of distinct loop configurations at any given degree grows much faster than on the hexagonal lattice. 

%%%%%%%%%%%%%%%%%%%%%%%%%%%%%%%%%%%%%%%%%
%%%%%%%%%%%%%%%%%%%%%%%%%%%%%%%%%%%%%%%%%

%%%%%%%%%%%%%%%%%%%%%%%%%%%%%%%%%%%%%%%%%
%%%%%%%%%%%%%%%%%%%%%%%%%%%%%%%%%%%%%%%%%

\section{Discussion} \label{sect:discussion}
A key result presented in this manuscript is a systematic way to improve on BP results for the contraction of tensor networks, turning BP from an uncontrolled approximation into a well-controlled approximation. This result would be directly useful in contexts where BP has been employed for the contraction of tensor networks, such as the simulation of quantum experiments\cite{QC11}.

Another direct application of these results could be for PEPS optimization algorithms. Given that a recent work\cite{BP13} established the equivalence of BP to the `simple update' strategy for PEPS, the series expansion could significantly enhance the accuracy of the `simple update' strategy while still remaining relatively computationally efficient. For instance, the genus-1 loop excitations on the hexagonal lattice, i.e.\ the $\delta_6$ and $\delta_{10}$ terms in Fig.~\ref{fig:free}, can be evaluated with $O(m^6)$ cost for PEPS of bond dimension $m$, and this cost could also potentially be reduced via approximation strategies like sparse diagonalization. Thus the dominant terms in the expansion are still feasible to compute even for PEPS with bond dims in excess of $m=100$, well beyond the computational limits of boundary MPS or corner transfer matrix methods for PEPS contractions\cite{PEPS5,PEPS6,PEPS7,PEPS7b,PEPS7c,PEPS8}. A general comparison of the computational efficiency of loop expansions versus MPS-based methods for iPEPS contractions remains beyond the scope of the present manuscript, given that such a comparison would be heavily dependent on the properties of the state under consideration. None-the-less, in Appendix.~\ref{sect:CTMRG} we demonstrate that certain (short-range correlated) PEPS states can indeed be contracted more efficiently using the loop expansion than with MPS-based methods.

We also expect the series expansions to be useful in the context of coarse-graining tensor algorithms for tensor networks, especially in relation to proposals that extend the tensor renormalization group\cite{TRG1} (TRG) by including mechanisms to identify and remove short-range loop correlations\cite{TRG2,TRG3,TRG4,TRG5}. Given that our results provide a general framework precisely for efficiently identifying loop correlations it is likely that they would find significant utility in this setting. 

Although our proposal worked well for the test cases considered, it is clear that there are also several potential ways that it could fail. Most obviously, application of the series expansion first requires finding a BP fixed point. It is unknown how difficult this task is for a PEPS representing the ground state of a local Hamiltonian or other physically relevant tensor networks. Even if a BP fixed point is found, the series expansion would not be expected to converge if the network under consideration possessed multiple degenerate, or almost degenerate, fixed points. In this scenario Eq.~\ref{eq:suppression} would fail; relative to any single BP fixed point there would exist high-degree configurations that still carry large weight (i.e related to a different BP fixed point). An example of this would be a PEPS encoding of a GHZ state, where a BP fixed point could realize either the $\ket{\uparrow\uparrow\uparrow\ldots}$ or $\ket{\downarrow\downarrow\downarrow\ldots}$ component of the network; relative to one of these fixed points the other component would represent a maximal-degree excitation with equal weight.  

Most existing tensor network algorithms are built on the twin pillars of tensor contractions and decompositions. The results presented in this manuscript, which prescribe a general procedure for resolving a complicated network into its fundamental components, could represent a third pillar for tensor network algorithms; one that could have interesting synergies with established routines. Exploring the ways that series expansions could compliment or enhance existing tensor network algorithms remains an interesting avenue for future research. 

% For instance, in evaluating a network $\mathcal T$, one could begin by expanding in terms of the BP vacuum, the short-loop excitations, and the remaining `residual'. Potentially, one could then apply a traditional (approximate) contraction strategy on the residual which, presumably, would be more amenable to truncations due to the separation from the short-loop excitations. Exploring the ways that series expansions could compliment or enhance existing tensor network algorithms remains an interesting avenue for future research. 

\begin{acknowledgments}
We thank AWS for supporting the quantum computing program. GKC acknowledges support from the US DOE, Office of Science, National Quantum Information Science Research Centers, Quantum Systems Accelerator (QSA), and a generous gift from AWS. We acknowledge funding provided by the Institute for Quantum Information and Matter, an NSF Physics Frontiers Center (NSF Grant PHY-2317110).
\end{acknowledgments}

\newpage
\appendix

\section{Exponential Suppression of High-degree Excitations} \label{sect:degree}
The utility of the proposed loop expansion is built on the premise that the weights of excitations (i.e. on top of the BP vacuum) are exponentially diminishing in their degree; this property allows us to accurately approximate a tensor network contraction using only a few low-degree excitations. In this appendix we attempt to justify this premise.

\begin{figure} [!t] %[!t!b]
\begin{center}
\includegraphics[width=7.0cm]{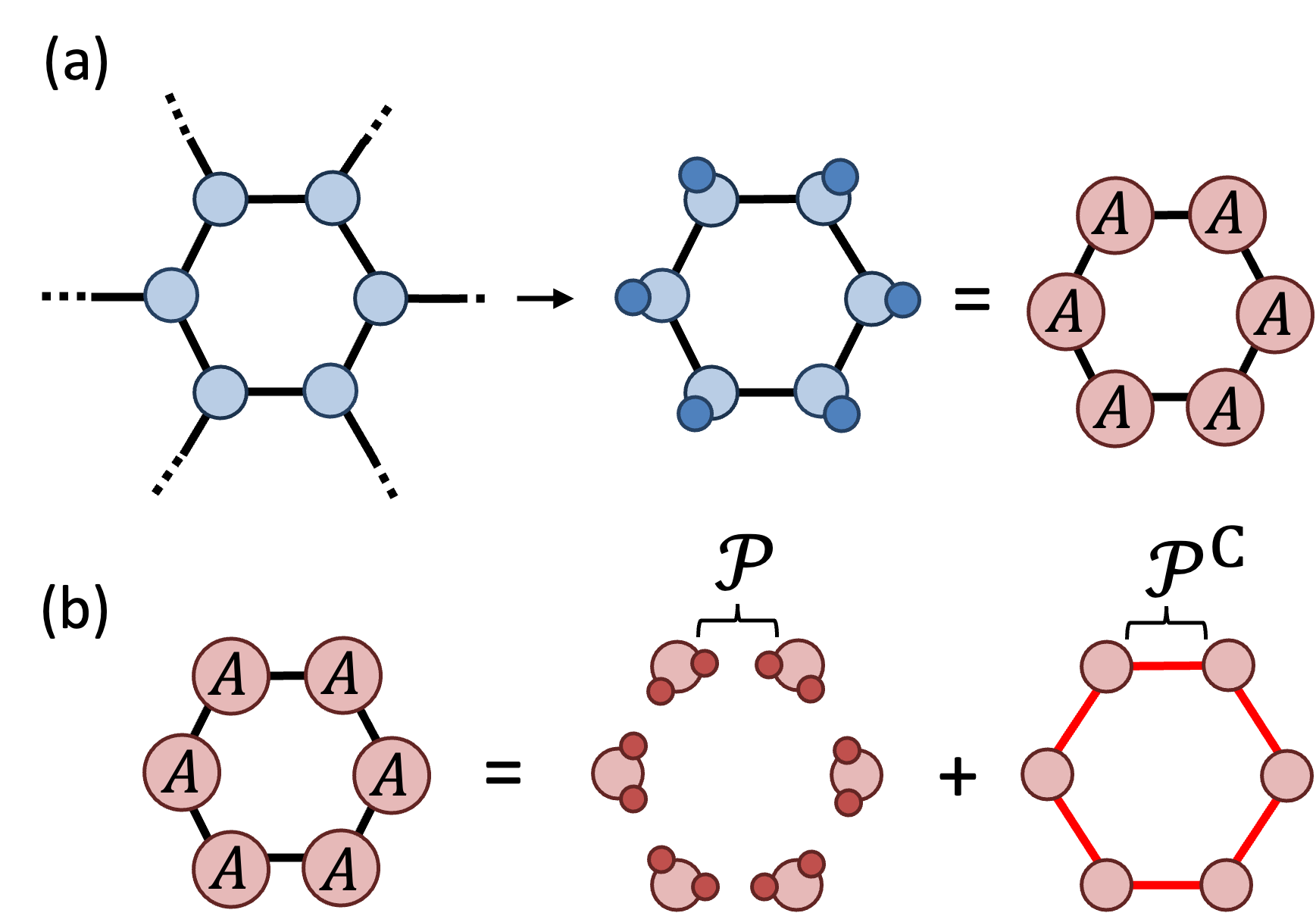}
\caption{(a) A loop of 6 tensors within an extended network is isolated by setting external edges with their fixed point messages and then absorbing the messages into the adjoining tensors. The resulting network is given as the trace of a product of matrices $A$, which are assumed to be identical due to symmetry. (b) The loop network is decomposed into the BP component (projecting each edge into the message subspace) plus the loop excitation (projecting each edge into the excited subspace), see also Eq.~\ref{eq:A1e1}.}
\label{fig:suppression}
\end{center}
\end{figure}

Let $\mathcal T$ be a finite tensor network containing at least a single closed loop. We assume that a BP fixed point has been found and that the tensors in network $\mathcal T$ have been normalized such that their vacuum contributions are unity (or equivalently that the Bethe free energy is zero). Finally, we also assume that the BP fixed point accurately approximates the network contraction; that the difference between the full network contraction $\mathcal Z(\mathcal T)$ and the BP approximation $\mathcal Z_\textrm{BP}$ is small. We now consider isolating a single length-$x$ loop of the network by absorbing the fixed point messages on all edges outside of the loop into their adjoining tensors, such that the reduced network is realized as the trace of a product of matrices as shown in the example of Fig.~\ref{fig:suppression}(a). For simplicity we assume that the matrices within the loop are copies of a single matrix $A$, which would only occur in practice if the original network possessed sufficient spatial symmetries, although the essence of our argument does not rely on this assumption. It follows that the network contraction can be written as $\mathcal Z(\mathcal{T}) = \tr\left\{ {A^x} \right\}$. The reduced network can be decomposed as the BP vs the non-BP component
\begin{equation}
\mathcal Z(\mathcal{T}) = \tr\left\{ {(A\mathcal{P})^x} \right\} + \tr\left\{ {(A\mathcal{P}^C)^x} \right\},\label{eq:A1e1}
\end{equation}
with $\mathcal P$ as the rank-$1$ projector onto message subspace and $\mathcal P^\textrm{C}$ its complement, see also Fig.~\ref{fig:suppression}(b). Note that $\mathcal Z_\textrm{BP} = \tr\left\{ {(A\mathcal{P})^x}\right\} = 1$ due to the normalization condition imposed. Given that BP fixed points are known to correspond to stationary points of the Bethe free energy, and that we have assumed our known fixed point accurately approximates the network contraction, it follows that the projector $\mathcal{P}$ onto message subspace should maximize $\mathcal Z_\textrm{BP}$. Thus $\mathcal{P}$ must equal the outer product of the dominant left/right eigenvectors of matrix $A$ (or, equivalently, the fixed point messages within the loop must by the left/right eigenvectors of $A$). Let $\{\lambda_0, \lambda_1, \lambda_2,\ldots\}$ be the eigenvalues of $A$ ordered with descending magnitude, where the normalization condition implies that $|\lambda_0|=1$. It follows that the weight $W(\delta_x)$ of the loop correction term can be written as
\begin{equation}
W(\delta_x) = \tr\left\{ {(A\mathcal{P}^\textrm{C})^x} \right\} = \sum_{r=1}^{d} \left( (\lambda_r)^{x} \right). \label{eq:A1e2}
\end{equation}
Let us now consider how Eq.~\ref{eq:A1e2} would scale in the limit of large length-$x$ (still assuming that the matrices within the loop are identical). Firstly, we should realize that the sub-leading eigenvalues should have magnitude strictly less than unity, i.e. $|\lambda_1| < 1$, otherwise the loop correction term would out-weight the BP term, violating the prior assumption that the BP fixed point accurately approximates the full network contraction $\mathcal Z(\mathcal T)$. In the large-$x$ limit Eq.~\ref{eq:A1e2} will be dominated be the largest magnitude eigenvalues, so if $\lambda_1$ of $A$ is $n$-fold degenerate it follows
\begin{equation}
W(\delta_x) \approx n (\lambda_1)^x \label{eq:A1e3}
\end{equation}
which may be rewritten as 
\begin{equation}
W(\delta_x) \approx e^{-kx} \label{eq:A1e4}
\end{equation}
with $k=-\log(\lambda_1) - (1/x)\log(n)$. Notice that $k$ tends to a positive constant with large-$x$, given that $\lambda_1 < 1$, such that the weight $W(\delta_x)$ tends to purely exponential decay in degree-$x$ as was posited Eq.~\ref{eq:suppression} of the main text.

%%%%%%%%%%%%%%%%%%%%%%%%%%%%%%%%%%%%%%%%%
%%%%%%%%%%%%%%%%%%%%%%%%%%%%%%%%%%%%%%%%%

Although our argument assumed, for simplicity, that the transfer matrices with any loop were identical this condition is not strictly necessary; it is only required that the transfer matrices have a gap between their dominant and sub-leading eigenvalues in order to realize (approximate) exponential decay. Finding a general characterisation of the networks that are expected to have a gap in the transfer matrix spectrum associated to any closed loop is likely a difficult task (and certainly beyond the scope of this manuscript). However, given that the loop series expansion is only expected to be viable for networks where the loop corrections are small in comparison to the BP approximation, it follows that the gap condition must hold in this circumstance. Finally we remark that our argument is only valid for excitations that realize genus-1 loops not, for instance, higher genus loops like the $\delta_{11}$ excitations shown in Fig.~\ref{fig:free}(d). None-the-less, in practice the scaling of Eq.~\ref{eq:A1e4} does seem to hold for higher genus loop excitations.

\section{Loop Excitations in the Thermodynamic Limit} \label{sect:counting}
In any finite tensor network it is possible, at least in principle, to enumerate over the different possible loop configurations, organize them by degree, and then evaluate the terms out to some desired degree. In contrast, in the thermodynamic limit, even counting the density of terms of some fixed degree may be an intractable combinatorial problem. To this end, we now discuss three different methods to approximate the loop corrections to the free energy density $f$, as defined in Eq.~\ref{eq:free}, of a homogeneous tensor network on a lattice in the thermodynamic limit (e.g. an iPEPS).

% We shall compute the contributions from configurations that have a single connected excitation, with all other tensor edges set in their BP ground state.

% We focus on the task of evaluating the free energy density $f_N$ per site, defined
% \begin{equation}
% f_N \equiv -\frac{\log(\mathcal Z_{M})}{2M}. \label{eq:appx1}
% \end{equation}

\subsection{Single-excitation approximation} \label{sect:single}
The first method that we propose to estimate the contribution of loop excitations to the free energy density involves counting only the configurations that involve only a single excitation on top of the BP vacuum. Let $\mathcal{T}_M$ be a homogeneous tensor network on a hexagonal lattice of $M$ plaquettes that, when contracted, evaluates to some scalar $\mathcal Z_{M}$. We begin by assuming that we have already found the BP fixed point of the network, and that the network has been normalized such that the BP vacuum is unity, $\mathcal Z_{\textrm{BP}} = 1$ (or equivalently that the Bethe free energy is zero). 

\begin{figure} [!t] %[!t!b]
\begin{center}
\includegraphics[width=8.5cm]{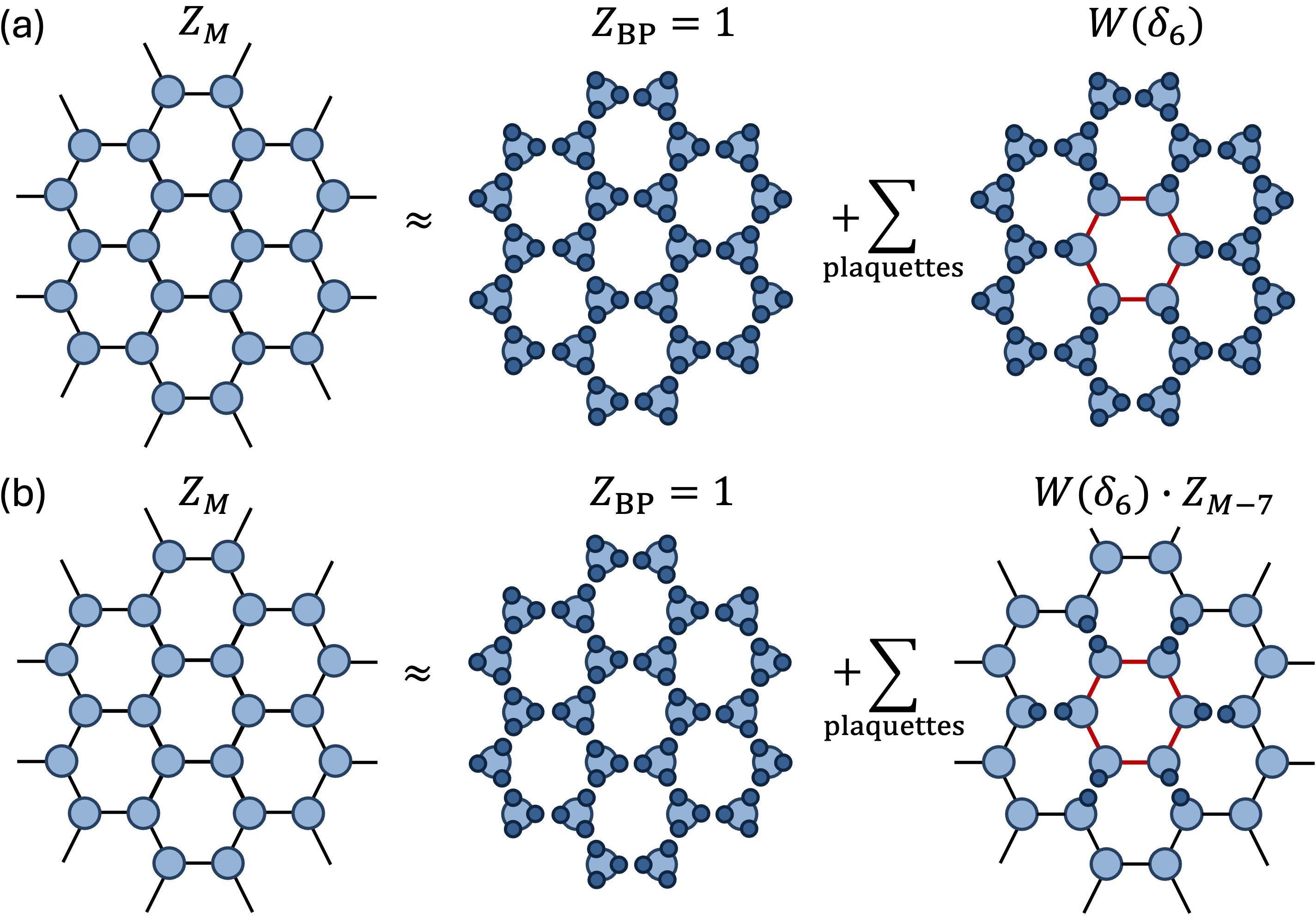}
\caption{(a) A hexagonal lattice tensor network is approximated via the \emph{single-excitation} approximation for the $\delta_6$ loop excitations. The tensor indices in each expansion term, except those which are part of the $\delta_6$ excitation, are set in the BP ground state. (b) The \emph{self-consistent completion} is used to expand the tensor network for the $\delta_6$ loop excitations. In each expansion term only the tensor indices directly connected to the $\delta_6$ excitation are set in the BP ground state.}
\label{fig:counts}
\end{center}
\end{figure}

We now consider evaluating the corrections arising from the $\delta_6$ excitations, see Fig.~\ref{fig:counts}(a), although our derivation is easily extendable to include arbitrary loop excitations. Given that there are $M$ distinct locations that the $\delta_6$ excitation could occupy, with each configuration having a weight of $W(\delta_6)$, it follows that the scalar $\mathcal Z_{M}$ of the network is 
\begin{equation}
\mathcal Z_{M} \approx 1 + M W(\delta_6) \label{eq:appx2}.
\end{equation}
The (per-site) free energy density can thus be approximated as 
\begin{align}
f &\approx \frac{-\log\left(1 + M W(\delta_6)\right)}{2M} \nonumber \\
& \approx -\frac{1}{2} W(\delta_6). \label{eq:appx3}
\end{align}
To get to the second line of working in Eq.~\ref{eq:appx3} we have used $\log(1+\epsilon) \approx \epsilon$ which requires $1/M \gg W(\delta_6)$ to be valid. That the single-excitation approximation is only valid for small $M$ makes sense; for larger $M$ lattices the contributions from configurations with double-excitations (or higher) will dominate as their multiplicity scales at least as $M^2$. However, given that the result from Eq.~\ref{eq:appx3} is independent of $M$, it may still be used as an approximation to the free energy density $f$ in the thermodynamic limit. Another way of interpreting this derivation is that we are using the single-excitation approximation to calculate the free energy density on only a small patch of the otherwise infinite system. We can easily generalize Eq.~\ref{eq:appx3} to include arbitrary excitations $\delta_x$,
\begin{equation}
f \approx \sum_{\{\delta_x\}} -L(\delta_x) W(\delta_x), \label{eq:appx4}
\end{equation}
where $L(\delta_x)$ is the number of locations per lattice site that the excitation $\delta_x$ can be placed.

%%%%%%%%%%%%%%%%%%%%%%%%%%%%%%%%%%%%%%%%%
%%%%%%%%%%%%%%%%%%%%%%%%%%%%%%%%%%%%%%%%%

\subsection{Self-consistent completion} \label{sect:multi}
The second method that we propose to estimate the contribution of loop excitations uses a self-consistent completion. For simplicity, we again focus on the $\delta_6$ excitations, noting that the derivation is easily extended to include arbitrary excitations. We now resolve the hexagonal network of $M$ plaquettes as the BP contribution $\mathcal Z_{\textrm{BP}}$ plus $M$ distinct contributions arising from the placement of a $\delta_6$ excitation on the lattice as shown Fig.~\ref{fig:counts}(b). However, in each contribution we freeze only edges connecting the excitation to their BP ground while leaving the other edges in the network free. It follows that each of these contributions has total weight $W(\delta_6) \tilde{\mathcal Z}_{M-7}$, where $\tilde{\mathcal Z}_{M-7}$ represents the scalar from contracting the remaining tensor network, which now consists of $(M-7)$ complete hexagonal plaquettes. Thus the scalar $\mathcal Z_{M}$ from the network contraction is approximated as
\begin{equation}
\mathcal Z_{M} \approx 1 + M W(\delta_6) \tilde{\mathcal Z}_{M-7}.  \label{eq:appx5}
\end{equation}

In order to make headway with Eq.~\ref{eq:appx5} another approximation is needed; we now assume that $\tilde{\mathcal Z}_{M-7}$ and ${\mathcal Z}_{M}$ have the same free energy density $f$, which implies
\begin{equation}
\tilde{\mathcal Z}_{M-7} \approx e^{7f} {\mathcal Z}_{M}. \label{eq:appx6}
\end{equation}
Note that this approximation for $\tilde{\mathcal Z}_{M-7}$ ignores boundary effects (i.e. resulting from fixing some of the network edges in their BP ground). Substitution of Eq.~\ref{eq:appx6} into Eq.~\ref{eq:appx5} gives
\begin{equation}
\mathcal Z_{M} \approx 1 + \mathcal Z_{M} \left(e^{7f} M W(\delta_6)\right), \label{eq:appx7}
\end{equation}
which can be rearranged as
\begin{equation}
\mathcal Z_{M} \approx \left(1 - e^{7f} M W(\delta_6)\right)^{-1}. \label{eq:appx8}
\end{equation}
Finally, the (per-site) free energy density $f$ can be expressed as
\begin{align}
f &\approx \frac{\log(1-e^{7f} M W(\delta_6))}{2M} \nonumber \\
& \approx -\frac{1}{2} W(\delta_6) e^{7f}, \label{eq:appx9}
\end{align}
where the approximation $\log(1+\epsilon) \approx \epsilon$ is used to obtain the second line of working. This formula is easily generalized to include arbitrary excitations $\delta_x$
\begin{equation}
f \approx \sum_{\{\delta_x\}} -L(\delta_x) W(\delta_x) e^{S(\delta_x) f}, \label{eq:appx10}
\end{equation}
with $L(\delta_x)$ as the number of locations per lattice site that the excitation can be placed and $S(\delta_x)$ as the number of hexagonal plaquettes that each excitation occupies. Notice that Eq.~\ref{eq:appx10} is identical to Eq.~\ref{eq:appx4} apart from the addition of the \emph{suppression factors} $\exp{(S(\delta_x) f)}$, which serve to approximate the contributions arising from configurations containing multiple excitations.

% which we refer to as \emph{bulk suppression factors} as they act to suppress excitations that occupy larger regions of the network. 

% The appearance of these factors in the multi-excitation derivation makes intuitive sense: excitations that occupy larger regions leave less `room' for other excitations to occur and are penalized accordingly (whereas this consideration is irrelevant if only single excitations are allowed).

Finally, we remark that, in general, it will not be possible to solve Eq.~\ref{eq:appx10} directly for (the loop corrections to) the free energy density $f$ as it appears on both sides of the equation. However, given that these corrections are assumed to be relatively weak (otherwise the series expansion itself would not be valid), we are able to use an iterative strategy to approximately solve for $f$. This strategy is as follows: first we set $f=0$ on the r.h.s of Eq.~\ref{eq:appx10} to compute a new value for $f$, before then feeding the new $f$ back into the r.h.s and repeating until $f$ is converged. In our benchmark calculations we observed that this strategy only required a few iterations in order to converge to high precision. 

% Before continuing our derivation it is informative to reflect on the advantages and disadvantages of the proposed approximation. A key advantage of this proposal is that it implicitly includes multi-excitations; when resolving a contribution containing a single $W(\delta_4)$ excitation the remaining network $\tilde{\mathcal T}_{N-4}$ could subsequently be resolved into a further sum of excitations. However, a significant disadvantage is that this approximation counts certain configurations multiple times, see Fig.~\ref{fig:counts}(c-d) for examples. While it could be possible to use a more sophisticated strategy that subtracts out contributions from multiply-counted configurations, here we simply ignore them as in practice they seem to have minimal impact on accuracy.

\subsection{Multi-excitation counting} \label{sect:multicount}
A third strategy to estimate the contribution of loop excitations is simply to explicitly count the products of distinct excitations \footnote{I. Lukin and A. Sotnikov, \emph{private communication.}}. Again we consider a network on a hexagonal lattice of $M$ plaquettes that evaluates to some scalar $\mathcal Z_{M}$. Here the loop corrections, when expanded out to the degree-10 terms, remain identical to those from the single-excitation approximation in Eq.~\ref{eq:appx4},
\begin{equation}
\mathcal Z_M = MW(\delta_6) + 3 NW(\delta_{10}). 
\end{equation}
However, the corrections from degree-12 excitations, $(\Delta \mathcal Z)_{12}$, now include the $\delta_{12}$ excitations in addition to the product of two degree-6 excitations $\delta_{6}$, 
\begin{equation}
(\Delta \mathcal Z)_{12} = 2 MW(\delta_{12}) + \frac{M(M-7)}{2} W(\delta_{6})^2, 
\end{equation}
where the factor $\frac{M(M-7)}{2}$ counts the number of distinct locations that the two degree-6 excitations can be placed. The corrections from the degree-14 excitations remain unchanged,
\begin{equation}
(\Delta \mathcal Z)_{14} = 12 NW(\delta_{14}),
\end{equation}
but the degree-16 excitations additionally include the product of $\delta_{6}$ and $\delta_{10}$ excitations,
\begin{equation}
(\Delta \mathcal Z)_{16} = 18 MW(\delta_{16}) + 3M(M-10) W(\delta_{6}) W(\delta_{10}).
\end{equation}
This methodology, which explicitly counts product excitations, can easily be extended out to arbitrary degree, e.g. where the degree-18 expansion would also include products of three $\delta_{6}$ excitations.

Taking the logarithm of $\mathcal Z_{M}$, the (per-site) free energy density $f$ is given as
\begin{align}
f = &-\frac{1}{2} W(\delta_6) -\frac{3}{2} W(\delta_{10}) -W(\delta_{12}) + \nonumber \\
& \frac{7}{4} W(\delta_{6})^2 -6 W(\delta_{14}) -9 W(\delta_{16}) + \nonumber\\
& 15W(\delta_{6}) W(\delta_{10}) + \ldots,
\end{align}
where it is noted that the terms of order $M^2$ canceled out in the logarithm expansion.

\subsection{Comparison} 
\begin{figure} [!t!h] %[!t!b]
\begin{center}
\includegraphics[width=8.5cm]{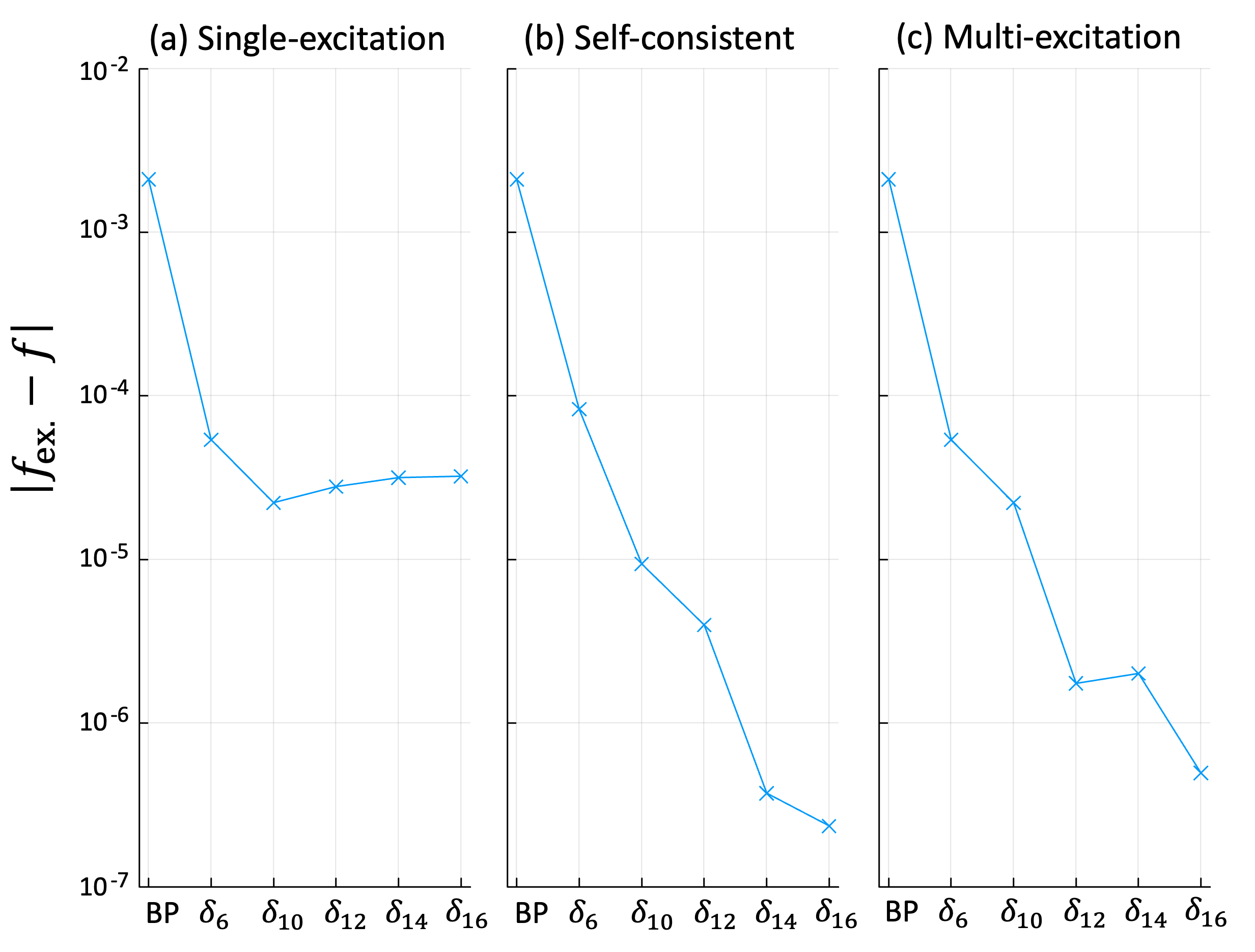}
\caption{Results comparing the accuracy of (a) the single-excitation approximation, (b) the self-consistent completion, and (c) the multi-excitation counting strategies for computing the free energy density $f$ of the hexagonal AKLT model when including all excitations $\delta_x$ up to degree $x$. While the single-excitation approximation saturates at a finite accuracy, the other strategies appear to converge to the exact result in the limit of a large degree expansion.}
\label{fig:compare}
\end{center}
\end{figure}

We have presented three approaches for using the loop expansion to estimate the free energy density on an infinite, homogeneous tensor network. A numerical comparison of these approaches is shown in Fig.~\ref{fig:compare} for loop corrections (expanded out to degree-16) in the hexagonal lattice AKLT model. While the expansion using the single-excitation approximation saturates at a finite accuracy, both the self-consistent completion and the multi-excitation approach appear to converge to the exact result as the degree of the expansion is increased. Similar results were also observed for the AKLT models on square and kagome lattices, and for randomly initialized PEPS tensor networks. 

While both the self-consistent completion and the multi-excitation counting are seen to give comparable accuracy, the numerical results of the main text were derived using the former approach. A reason for this is that the self-consistent completion is easier to apply to the case of computing transfer and density matrices, given that the same suppression factors $\exp{(S(\delta_x) f)}$ can be used as for the free energy density calculation. 

% These results demonstrate that the bulk suppression factors $\exp{(S(\delta_x) f)}$, which appear only in Eq.~\ref{eq:appx10} derived using the multi-excitation approximation, are vital to converge the loop expansions to high accuracy.

% While both approximations show similar accuracy when the expansion includes only the $\delta_6$ excitation, the accuracy of single-excitation approximation shows minimal improvement as higher-degree terms are included in the expansion. In contrast, the multi-excitation approximation appears to converge to the exact result as the expansion is taken to include higher-degree terms. 

%%%%%%%%%%%%%%%%%%%%%%%%%%%%%%%%%%%%%%%%%
%%%%%%%%%%%%%%%%%%%%%%%%%%%%%%%%%%%%%%%%%

\section{Kagome and Square Lattices} \label{sect:kagome}
\begin{figure} [!thb] %[!t!b]
\begin{center}
\includegraphics[width=8.0cm]{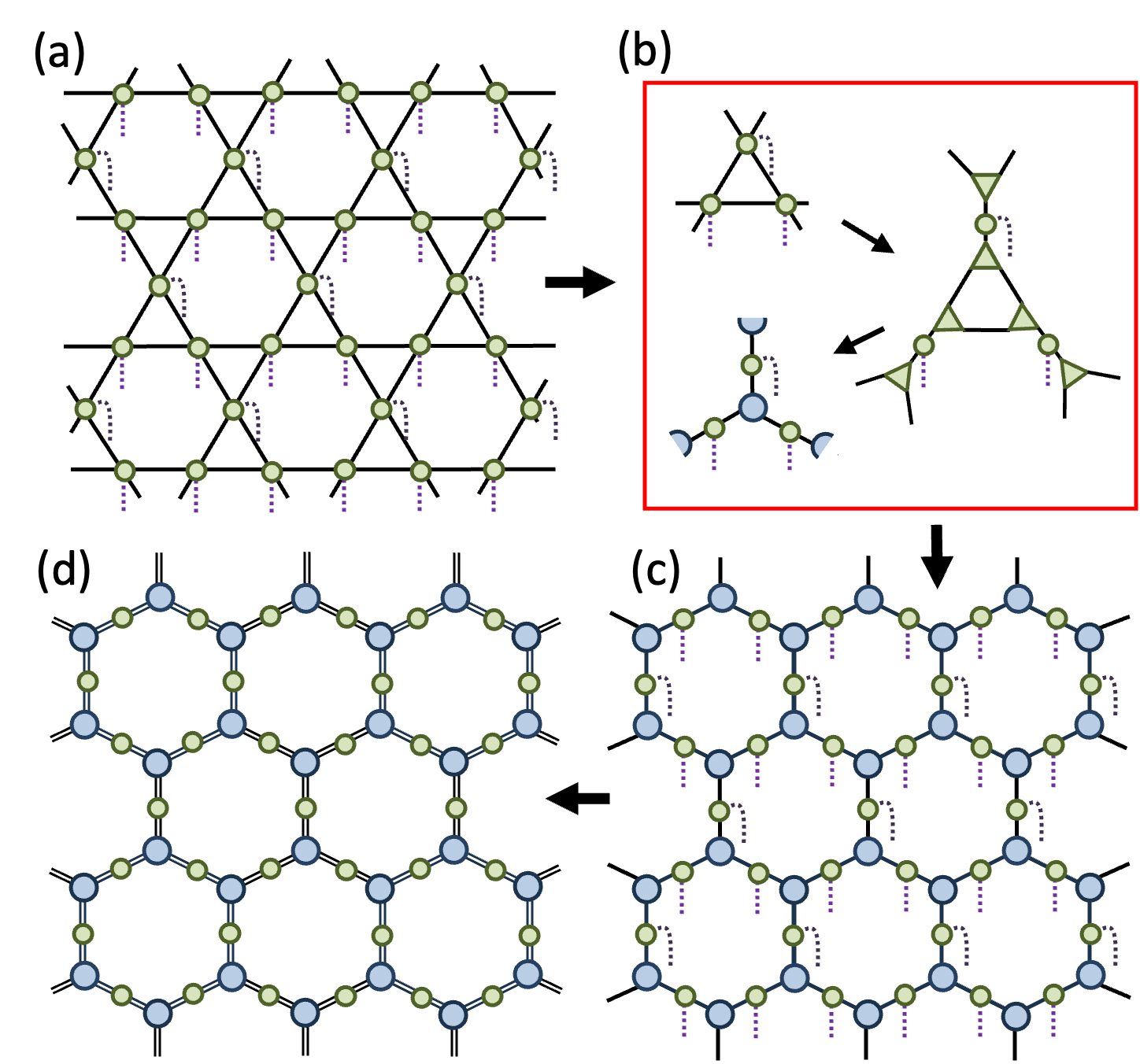}
\caption{(a) A kagome lattice iPEPS representing quantum state $\ket{\psi}$. (b) Each PEPS tensor is decomposed into a product of three 3-index tensors via singular value decompositions, then the tensors surrounding each triangle of the kagome lattice are contracted into a single tensor. (c) A modified PEPS representation of the quantum state $\ket{\psi}$ is obtained. (d) The square norm $\braket{\psi}{\psi}$ now evaluates to a hexagonal lattice tensor network.}
\label{fig:kagome}
\end{center}
\end{figure}

In this appendix we provide some of the details of the benchmark results provided for iPEPS on kagome and square lattices. An iPEPS for the kagome lattice is shown in Fig.~\ref{fig:kagome}(a). Although we could perform the series expansion on the (square norm) of this network directly, we instead found it preferable to perform some preliminary restructuring. The reasoning for this was to remove the triangular loops from the network which are otherwise problematic since (i) the weight of the excitations around the triangular loops are likely to be large (since the loops are short) and (ii) they generate a multitude of low-degree terms that would need to be included in the expansion. The restructuring is depicted in Fig.~\ref{fig:kagome}(b), where each (5-index) PEPS tensor is first decomposed as a product of three 3-index tensors, then the tensors surrounding each triangle of the kagome lattice are subsequently contracted into a single tensor (located at the center of each kagome lattice triangle). These manipulations result in a PEPS with a hexagonal lattice geometry but where the physical indices are located on tensors set on the edges of the hexagonal lattice as shown in Fig.~\ref{fig:kagome}(c). Upon forming the square norm $\braket{\psi}{\psi}$ a purely hexagonal lattice tensor network is obtained, as shown in Fig.~\ref{fig:kagome}(d), which can be treated with the same expansion as described for the hexagonal lattice PEPS. Note that the manipulations performed in Fig.~\ref{fig:kagome}(b) may require an increase in PEPS bond dimension from $m\to m^2$, although in many cases the new bond dimension could be truncated without incurring significant error. This example of manipulating the kagome lattice into a hexagonal lattice highlights an interesting point: the accuracy of the series expansion can depend on the particulars of the tensor network representation of a quantum state $\ket{\psi}$, rather than solely on the properties of $\ket{\psi}$ itself (e.g.\ entanglement and correlations). 

\begin{figure} [!thb] %[!t!b]
\begin{center}
\includegraphics[width=6.0cm]{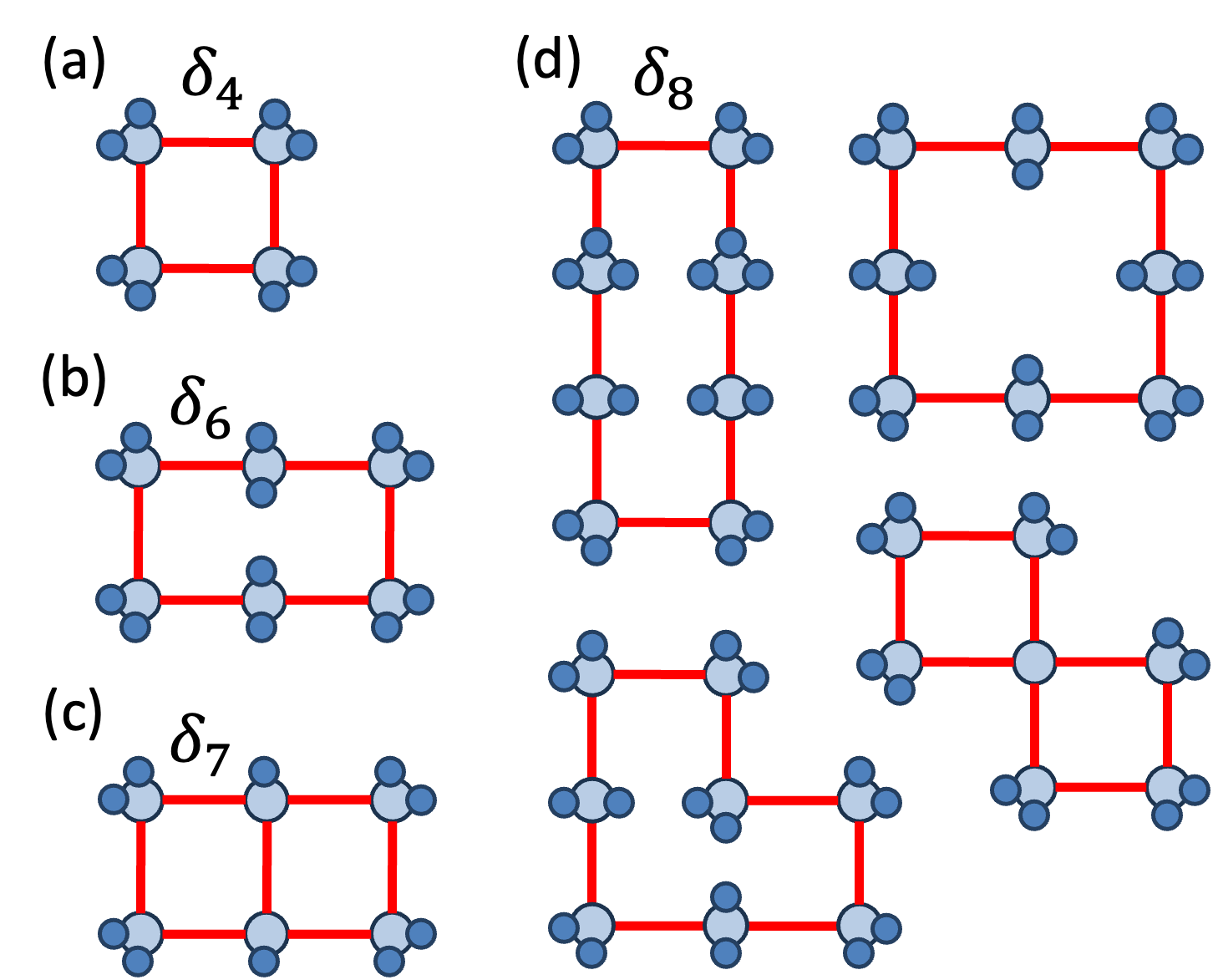}
\caption{Loop correction terms for a square lattice tensor network with a 1-site unit cell. (a) The $4^\textrm{th}$ degree term. (b) One (out of two) of the $6^\textrm{th}$ degree terms. One (out of two) of the $7^\textrm{th}$ degree terms. Four (out of nine) of the $8^\textrm{th}$ degree terms.}
\label{fig:square}
\end{center}
\end{figure}

The benchmark results of Sect.~\ref{sect:benchmark} also explore contraction of PEPS on the square lattice, both for the ground-state of the AKLT model as well as for randomly initialized networks. The square lattice results were computed similarly to the methods prescribed for the hexagonal lattice, both in terms of evaluating the series expansion and of evaluating the numerically exact results using boundary MPS. Several of the loop correction terms are shown in Fig.~\ref{fig:square}, where it can be seen that the number of distinct loop excitations of degree-$x$ scales more quickly as a function of $x$ than was seen with the hexagonal lattice. 

%%%%%%%%%%%%%%%%%%%%%%%%%%%%%%%%%%%%%%%%%
%%%%%%%%%%%%%%%%%%%%%%%%%%%%%%%%%%%%%%%%%

%%%%%%%%%%%%%%%%%%%%%%%%%%%%%%%%%%%%%%%%%
%%%%%%%%%%%%%%%%%%%%%%%%%%%%%%%%%%%%%%%%%

%%%%%%%%%%%%%%%%%%%%%%%%%%%%%%%%%%%%%%%%%
\begin{figure} [!t!h] %[!t!b]
\begin{center}
\includegraphics[width=7.5cm]{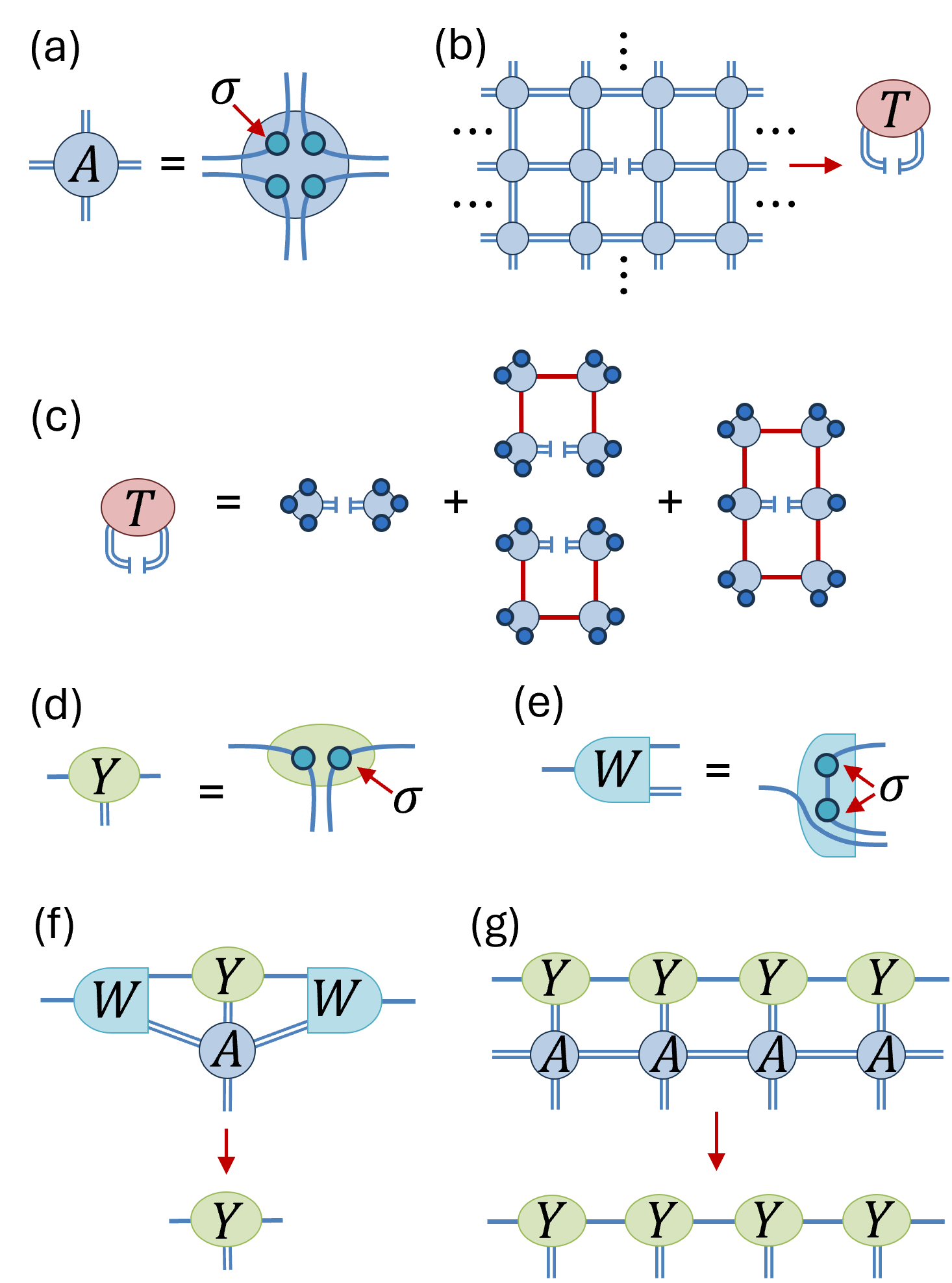}
\caption{(a) Definition of a corner double line (CDL) tensor. (b) A transfer matrix $T$ formed by opening a single index from an infinite network of CDL tensors. (c) The transfer matrix $T$ is exactly computed from the BP vacuum plus the single and double loop corrections. (d) Definition tensor $Y$. (e) Definition of isometry $W$. (f-g) A boundary MPS, formed from copies of $Y$, is seen to be an exact fixed point of the boundary MPS contraction when implemented in conjunction with isometries $W$.}
\label{fig:CDL}
\end{center}
\end{figure}
%%%%%%%%%%%%%%%%%%%%%%%%%%%%%%%%%%%%%%%%%

\section{Contraction Efficiency for PEPS versus MPS-based Approaches} \label{sect:CTMRG}
In the main text we demonstrated the use of the loop expansion to approximately contract PEPS tensor networks, although we did not attempt to quantify the efficiency (in terms of accuracy versus computational cost) versus established PEPS contraction methods such as boundary MPS or corner transfer matrix renormalization group (CTMRG). One reason for this was that a proper comparison would require a thorough and dedicated investigation as the relative accuracy is likely highly variable on the test model under consideration, and also that the cost scaling of MPS-based methods can depend on the finer details of their implementation. Given that our aim was to present the loop expansion as a general method to approximate arbitrary tensor network contractions, with the application to contract PEPS only as a single example, we felt such a comparison was beyond the scope of the present work. None-the-less, in this appendix we present an example that demonstrates that there exist $2D$ networks that are demonstrably more efficient to contract with a loop expansion than with previous MPS-based approaches.

Consider a homogeneous square-lattice tensor network of bond dimension $D$ formed from corner double-line (CDL) tensors, themselves composed of corner-sharing entangled pairs mediated by matrices $\sigma$ of dimension $\sqrt D$ as shown in Fig.~\ref{fig:CDL}(a). We assume that the $\sigma$ matrices are diagonal and contain an exponentially decaying weights,
\begin{equation}
\sigma_{ij} = \delta_{ij} \exp(-\alpha i) \label{eq:CDL}
\end{equation}
with $\alpha$ a positive constant. We focus on the task of computing the transfer matrix $T$ formed by cutting open a single index from the infinite network as shown in Fig.~\ref{fig:CDL}(b). We begin by computing $T$ using the loop series expansion, where computing the series expansion out to the double-plaquette term, see Fig.~\ref{fig:CDL}(c), is found to give the exact result. The cost of evaluating the double-plaquette correction is $O(D^4)$, although this cost reduces to $O(D^3)$ if only the single-plaquette loop corrections are considered, which may still provide a good approximation to $T$.

%%%%%%%%%%%%%%%%%%%%%%%%%%%%%%%%%%%%%%%%%
\begin{figure} [!t!h] %[!t!b]
\begin{center}
\includegraphics[width=8.0cm]{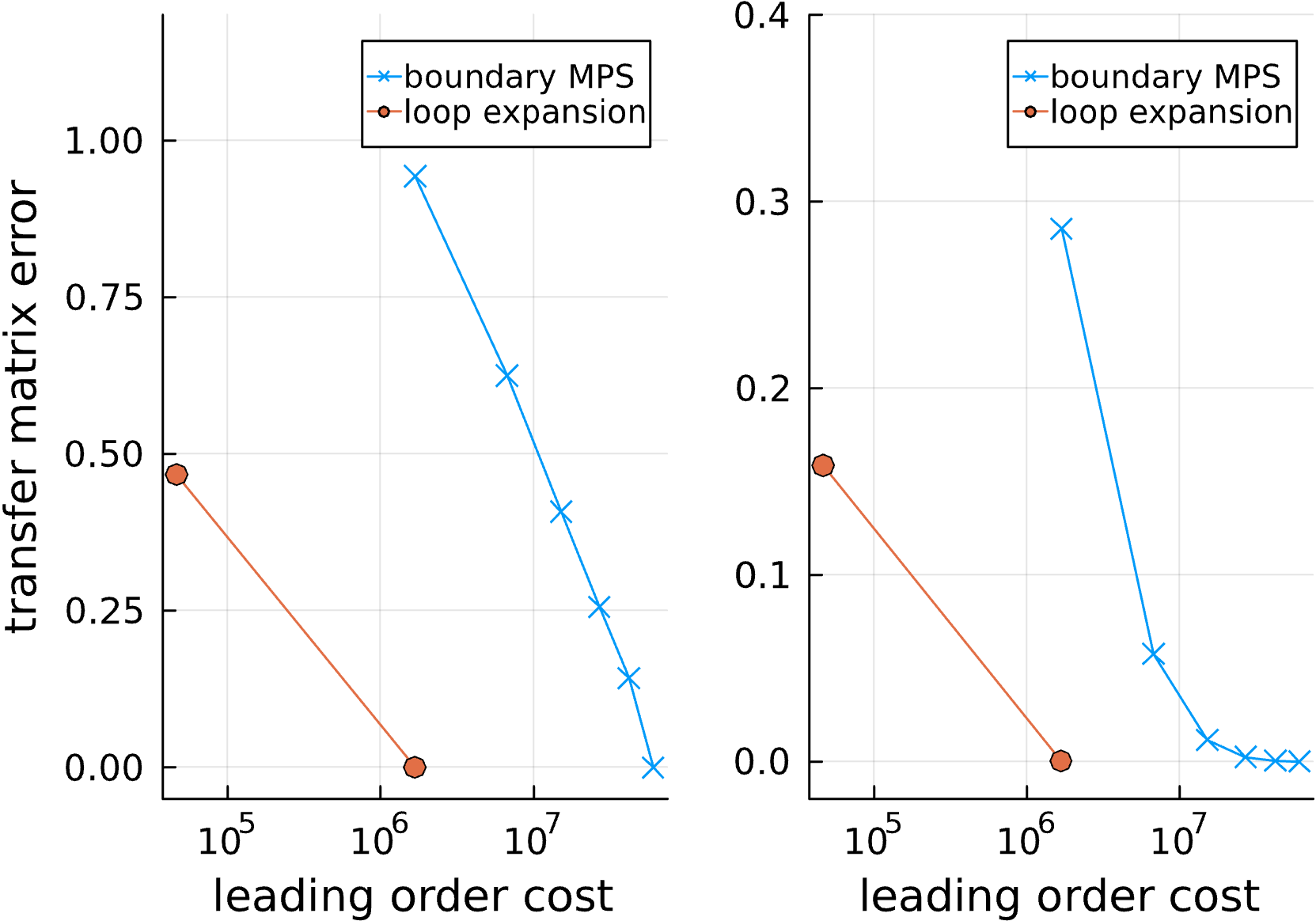}
\caption{Leading order computational cost versus accuracy for the error in the transfer matrix, $|| T - T_\textrm{exact} ||$, from a network of CDL tensors of dimension $D=36$ for (a) $\alpha=0.1$ and (b) $\alpha=0.4$ as defined in Eq.~\ref{eq:CDL}. Compared are bond dimensions $m=1,2,\ldots,6$ for the boundary MPS versus the loop expansion with either single or double loop corrections. The results demonstrate that the loop expansion is the more efficient approach: it produces higher accuracy for lower computational cost.}
\label{fig:efficiency}
\end{center}
\end{figure}
%%%%%%%%%%%%%%%%%%%%%%%%%%%%%%%%%%%%%%%%%

We now consider calculating $T$ through a boundary MPS approach\cite{PEPS5}. Alternatively, a corner transfer matrix approach\cite{PEPS6} could be used, although this is found to be equivalent in efficiency for the network under consideration. It can be seen that an MPS formed from tensors $Y$ as defined in Fig.~\ref{fig:CDL}(d), in conjunction with isometries $W$ as defined in Fig.~\ref{fig:CDL}(d), produces an exact fixed-point boundary MPS as shown in Figs.~\ref{fig:CDL}(f-g). The transfer matrix $T$ can then be evaluated from (two copies of) the boundary MPS using standard methodology. If the bond dimension $m$ of the boundary MPS is fixed at $m=\sqrt{D}$ then the contraction for $T$ is exact, although accurate approximations to $T$ can be obtained with smaller $m$. The leading order cost of contracting for $T$ using the boundary MPS approach is known to scale as $O(D^4 m^2)$. 

In Fig.~\ref{fig:efficiency} we compare the efficiency of the loop expansion versus the boundary MPS approach (in terms of the error $|| T_{\textrm exact} - T_{\textrm{approx}}||$ versus the leading order computational cost) for several values of $\alpha$ in Eq.~\ref{eq:CDL}. It can be seen that the loop expansion is the more efficient approach for contracting the CDL network: it produces more accurate results with less computational effort over the values of $\alpha$ considered. Although the CDL tensor network only has a simple entanglement structure, this example none-the-less demonstrates that the loop expansion may offer a more efficient contraction than traditional MPS-based approaches for contraction of certain $2D$ tensor networks.

%%%%%%%%%%%%%%%%%%%%%%%%%%%%%%%%%%%%%%%%%%%%%%%%%%%%%%%%%%%%%%%%%%%%%%%%%

\end{document}